\documentclass[10pt,journal,letterpaper]{IEEEtran}
\usepackage[colorlinks]{hyperref}
\usepackage{amsmath,bm} 
\usepackage[linesnumbered,ruled]{algorithm2e}
\usepackage{tabularx}
\usepackage{amssymb}
\usepackage{amsthm}

\usepackage{graphicx}
\usepackage{soul}
\usepackage{cite}
\usepackage[square,numbers]{natbib}
\usepackage{subfigure}
\usepackage{amsthm}
\usepackage{amsmath}
\usepackage{cleveref}
\usepackage{caption}
\usepackage{epstopdf}
\usepackage{indentfirst}
\usepackage{float}
\usepackage{mathrsfs}
\usepackage{epstopdf}

\usepackage[dvipsnames]{xcolor}
\begin{document}
\bstctlcite{IEEEexample:BSTcontrol}
\title{Analysis of Temporal Robustness in Massive Machine Type Communications}
	\author{Debjani Goswami$^{1}$, Merim Dzaferagic$^{2}$, Harun Siljak$^{2}$, Suvra Sekhar Das$^{1}$, and Nicola Marchetti$^{2}$
\\ G. S. Sanyal School of Telecommunications$^{1}$, Indian Institute of Technology Kharagpur, India.\\
	Trinity College$^{2}$, Dublin\\
	Email: debjani.ami.89@gmail.com,   DZAFERAM@tcd.ie, harun.siljak@tcd.ie, suvra@gssst.iitkgp.ernet.in,
	nicola.marchetti@tcd.ie}
	\maketitle
	\begin{abstract}
The evolution of fifth generation (5G) networks needs to support the latest use cases, which demand robust network connectivity for the collaborative performance of the network agents, like multi-robot systems and vehicle to anything (V2X) communication. Unfortunately, the user device's limited communication range and battery constraint confirm the unfitness of known robustness metrics suggested for fixed networks, when applied to time-switching communication graphs.
Furthermore, the calculation of most of the existing robustness metrics involves non-deterministic polynomial-time complexity, and hence are best-fitted only for small networks. Despite a large volume of works, the complete analysis of a \textit{low-complexity} temporal robustness metric for a communication network is absent in the literature, and the present work aims to fill this gap. More in detail, our work provides a stochastic analysis of network robustness for a massive machine type communication (mMTC) network. The numerical investigation corroborates the exactness of the proposed analytical framework for temporal robustness metric. Along with studying the impact on network robustness of various system parameters, such as cluster head (CH) probability, power threshold value, network size, and node failure probability, we justify the observed trend of numerical results probabilistically. \\\\
\IEEEpeerreviewmaketitle
\vspace{-7 mm}
\end{abstract}
\section{Introduction}
Due to the fast pace of technological advancement, current 5G systems need to support various brand-new network service requirements. In this context, the international telecommunication union (ITU) has categorized all 5G use cases into three broad service classes, i.e., enhanced mobile broadband (eMBB), ultra-reliable low latency communication (URLLC), and mMTC \cite{Service_Classification}. The use of multiple antennas at both the
transmitter and receiver sides, densifying networks with many
small base stations, and exploiting non-orthogonality in the
access scheme \cite{NOMA_MY}, are identified as the forefront techniques to boost the spectral efficiency (SE) in 5G systems.
Along with the three-fold SE requirement compared to 4G (eMBB service requirement), 5G also needs to support one million connections per square kilometer for the mMTC use case. Moreover, URLLC applications demand a 99.999\% reliability for transferring information within a one millisecond (ms) user plane latency. Two of the services mentioned above, i.e., URLLC, and mMTC, exploit the internet of things (IoT) networks formed by a group of wirelessly interconnected MTC devices (MTCDs). The small battery constraints of sensor nodes necessitate energy-efficient communication for the IoT use cases as the wireless sensor nodes act as MTCDs. Furthermore, many diverse application scenarios of 5G systems, e.g., multi-robot systems \cite{Multi-Robot}, unmanned aerial vehicles (UAV) \cite{UAV}, on-road-sensor networks (ORSN) \cite{ORSN}, air transportation system \cite{Air_Transportation}, smart grids, vehicular networks, all require a robust network to exploit the collaborative performance of sensor nodes. In this context, to enhance the connection robustness, 3GPP Rel 13 \cite{3gpp_37.340} has introduced the concept of dual connectivity that allows a single user to access two serving nodes or cell groups simultaneously. Unlike the infrastructure network, the limited communication range caused by the low transmitted power of sensor nodes limits the link connectivity in the network. Moreover, the random fading effect and the high mobility in vehicular networks results in communication link variation over both the nodes and time instants.   
Furthermore, as sensor nodes are operated automatically without human intervention, a frequent recharge of battery or replacement is difficult for sensor node-assisted applications. Consequently, a fast battery draining and physical damage in sensor nodes introduce random node failure, thereby significantly hampering networks' standard operation. These fundamental causes of network disruption raise the gateway failure robustness requirement for mMTC \cite{Robusness_Tech_MTC}, and the need for mobility robustness in the vehicle to anything (V2X) services \cite{Robusness_Tech_V2X} in 5G systems. 
Additionally, due to the recent paradigm shift from centralized to distributed, the robustness metric significantly evaluates any distributed algorithm's efficiency. For example, achieving a standard time scale within the network agents for exploiting the best benefit from the SE boosting techniques, requires a robust distributed timing synchronization solution for large networks {\cite{Time_Sync_My}}. The discussion above sheds light on the fact that irrespective of the service classes of 5G systems, robustness is one of the most desirable features for any communication network. In particular, the robustness metric captures the ability of a network to continue the normal operations in the presence of various disruptions and challenges. Although network robustness has gained extensive research attention in the last two decades, the above-mentioned new application scenarios and their unique service requirements necessitate a re-investigation of this domain. Hence, we now turn to discuss the motivation of our work in the context of existing network robustness metrics and their limitations in communication networks.

The earliest works exploit graph topology for quantifying  the network robustness, namely, the vertex (edge) connectivity \cite{Vertex_Connectivity}, heterogeneity \cite{heterogeneity}, clustering coefficient \cite{Clustering_Coefficient}, closeness \cite{closeness}, betweenness \cite{closeness}, diameter \cite{diameter}, algebraic connectivity \cite{Alge_Connec}, network criticality \cite{Network_Criticality}, effective graph resistance \cite{effect_graph_res}, number of spanning trees \cite{Hal_Spanning_Tree}, largest subgraph \cite{Giant_Components}, etc. Later works introduce some advanced metrics of network robustness, like, toughness \cite{Toughness}, scattering
number \cite{Scatting_number}, tenacity \cite{Tenacity}, which quantify both the cost of damage and the degree of impairment in networks \cite{Natural_Connectivity}. Unfortunately, as mentioned earlier, most of the existing robustness metric exhibit a non-deterministic polynomial-time complete (NP-complete) complexity, limiting the application of these metrics for large networks. Furthermore, the vertex or edge connectivity only partly reflects the disruption-withstanding capability of a network. Moreover, the algebraic connectivity for all disconnected graphs is zero \cite{Algebric_Connectivity_Survey}. The above limitations restrict the application of these metrics in many real networks. Furthermore, all the metrics above are suggested for fixed infrastructure networks, like the internet.
However, due to the random topology changes caused by fading, mobility, and energy constraints,  the robustness analysis becomes challenging \cite{Connectivity_Changes} for communication networks,
 as the network becomes a dynamic entity, driving the search for a candidate temporal robustness metric for switching networks.
 The first attempt for measuring the robustness aspect for time-varying networks is introduced in  \cite{Scellato_1}, where the authors express the temporal efficiency in the observation time interval $[t_{1},t_{2}]$ as the mean inverse temporal distances expecting over all pairs of nodes. Later on, the same authors of \cite{Scellato_1} extend their work in \cite{Temp_1}, where they define a novel temporal robustness metric to measure the degree of network disruption due to different attacking strategies. This work confirms that intelligent attacks have a more substantial influence on temporal connectivity than random attacks in the networks. Motivated by the fact above, a semi-supervised spatio-temporal deep learning-based intrusion detection scheme is proposed in \cite{Semi-supervised}. Authors in \cite{ATN} investigate the temporal robustness metric for an air transport network, where they confirm that although the temporal variation cannot affect the traditional giant component \cite{Giant_Components}, it has a more substantial effect on efficiency as described in \cite{Scellato_1}. Moreover, the authors report that the attacks based on the betweenness significantly degrade the temporal efficiency, or in other words, damage the network most. In the initial stage of the investigation, the temporal robustness evaluation follows the splitting of time switching graphs into a collection of static graphs, each for a single time instance and analyzing them one by one to produce the final result. A fast algorithm to evaluate the temporal robustness exploiting a single-stage computation is suggested by \cite{Algorith_TR}, thereby experiencing a lower time complexity than the previous approach. The authors in \cite{Railway_Transportation} apply temporal network theory for investigating the temporal features of railway transportation service networks, confirming the heterogeneous influence of different failing edges in the network.
The work in \cite{Nature_TR} defines the temporal robustness metric as the ratio
 between the average degree of nodes after and before the network disruption.
Moreover, work in \cite{Swarm_Systems} has introduced a new reliability measurement and named it cooperation reliability (CR) which captures both system integrity and motion consensus for swarm systems. This work confirms that the reliability of the swarm system varies with the density of the network agents. 
In the recent investigation in \cite{Recent_IoT}, the authors exploit various graphs to model the IoT network and provide the theoretical foundations of network criticality as a function of different network parameters.
Although most of the works mentioned above come up with some foremost insights, none of the works described in \cite{Scellato_1}-\cite{Recent_IoT} consider the impact of transmission range although it has a significant influence on how well a network can combat disruption. Although \cite{Numerical_Commu_Range} addresses the effect of random node distribution and node transmission range on network robustness in wireless sensor networks (WSNs) numerically, the mathematical analysis is beyond the scope of that work. 
To bridge the gap of complete investigation on temporal robustness, the authors of \cite{Temporal_Robust} present a markov-based temporal network model for analyzing the proposed temporal robustness metric. However, that work assumes the presence of a complete graph, which is not always ensured, especially in large networks.
  The above discussion reveals that despite the large volume of works towards measuring network robustness, a detailed analysis of a suitable temporal robustness metric for a communication network is still missing, which we aim to investigate in this work.\\
The contributions of our work can be summarized as follows:\\
$\bullet$ We provide a stochastic analysis of the temporal robustness metric. Moreover, we validate the correctness of our numerical investigation measuring the temporal robustness metric, by comparing with a prior temporal robustness metric described in {\cite{Nature_TR}}. The good match of analytical and numerical analysis confirms the exactness of the derived analytical expression.\\ 
$\bullet$ We investigate the behaviour of network robustness versus different system parameters, i.e., CH probability and the number of nodes in the network for different values of power threshold. Moreover, the observed trends are justified in a probabilistic manner.\\
$\bullet$ Unlike the other existing robustness metrics that experience NP-time complexity, our proposed metric has a complexity of $O(N^3)$ for the derived exact expression and $O(N^2)$ for its approximated form, making it suitable for large networks.
\section{System Model}
\begin{figure}
    \centering
    \includegraphics[width=1\linewidth]{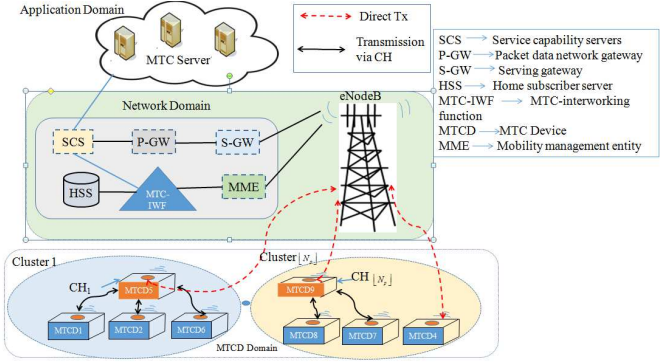}
     \caption{Cluster based network architecture for massive machine type communication services \cite{mMTC_Architecture}.}
     \label{Cluster_mMTC}
     \end{figure}

 The previous section confirms that a significant performance degradation caused by the limited communication range and tight energy constraint of sensor nodes in mMTC services necessitates gateway robustness in the mMTC network. Our work is motivated by the fact that the influence of these two aspects makes the robustness analysis challenging \cite{Connectivity_Changes} in mMTC networks. Furthermore, the massive connectivity in mMTC services results in a significant increase in the signaling overhead, thereby contributing to network congestion. One possible way to mitigate the congestion in the evolved NodeB (eNB)’s access channel \cite{Clustering1}-\cite{Clustering3} and exploiting the energy-efficient communication \cite{Clustering}-\cite{Clustering_2} is to group the MTCDs into smaller clusters. Motivated by that, we adopt the system model of a cluster-based mMTC network {\cite{Clustering}} to study the robustness of the network, as shown in Fig. \ref{Cluster_mMTC}. 
In particular, this investigation considers a fixed network size of $N$ where we exclude the possibility of adding new nodes in the network. We assume that all $N$ nodes are uniformly distributed inside a square grid within the range $[-a, a]$, i.e., the area of the grid is $A = 4a^2$, and the density of the nodes is $\lambda = N/A$. Moreover, we assume the base station (BS) is placed in the center of the square grid (i.e., in the (0,0) coordinate). 
\begin{figure}[H]
	\centering
	\subfigure[]{\label{System Model}\includegraphics[width=0.4\textwidth]{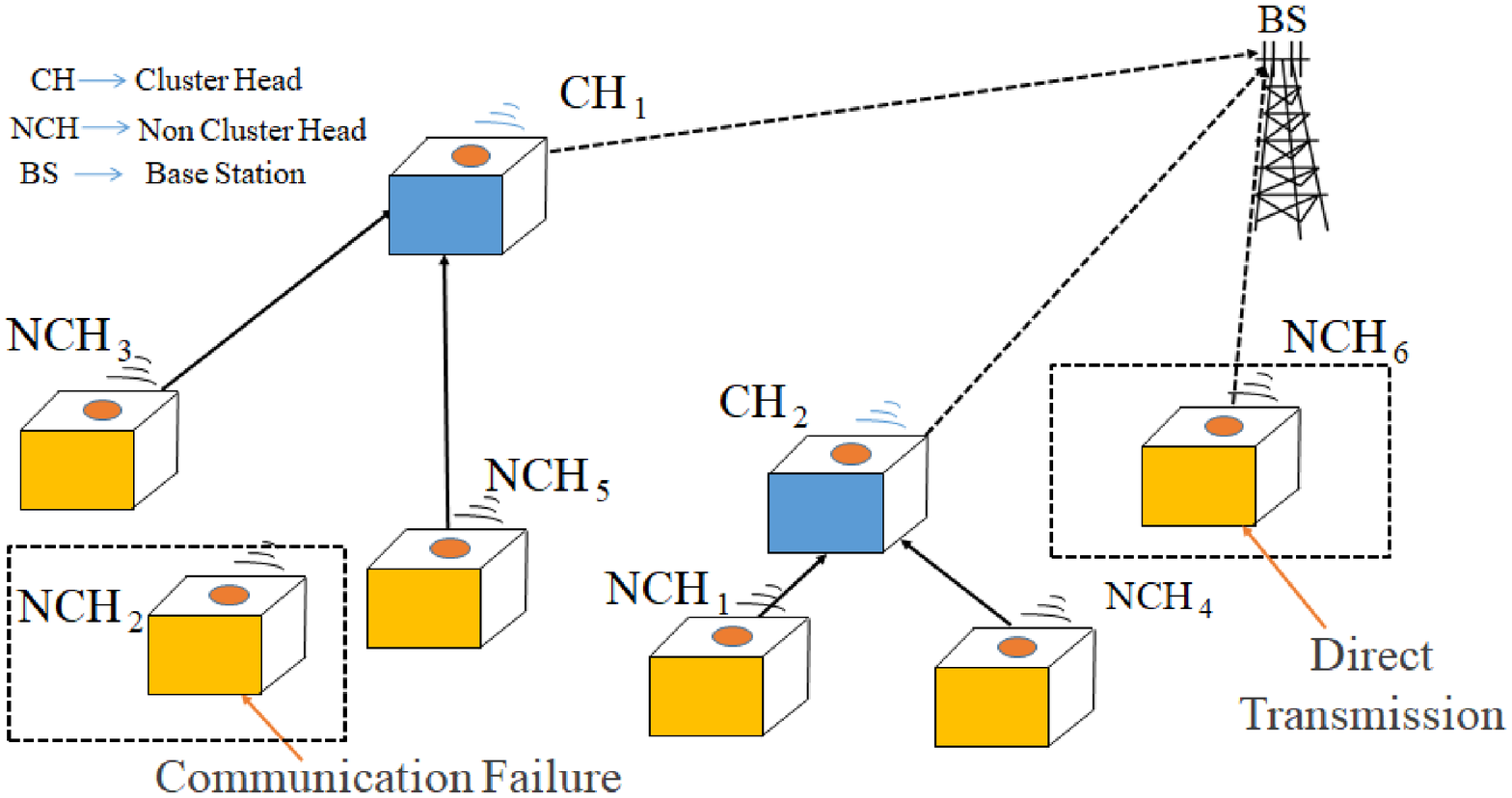}}\hfill
	\subfigure[]{\label{Association Strategy}\includegraphics[width=0.4\textwidth]{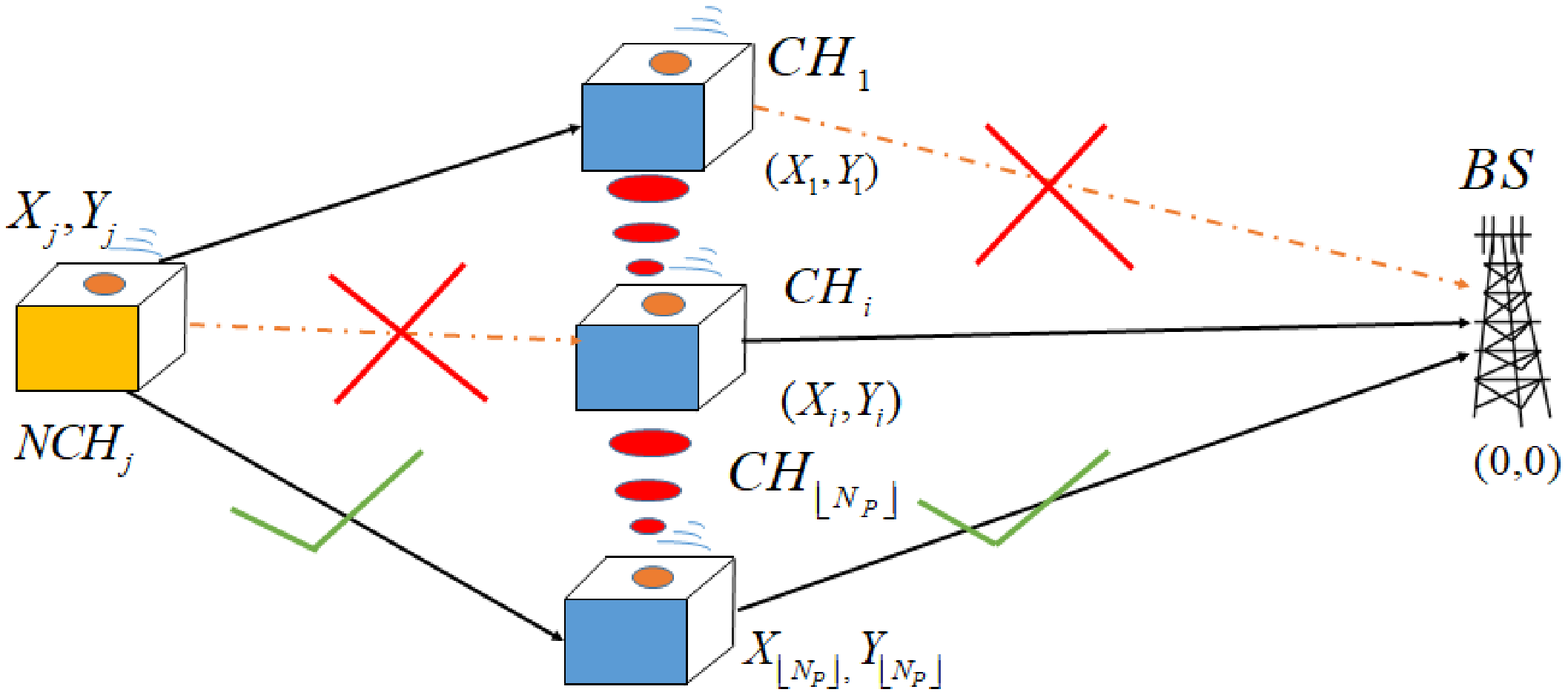}}\hfill
	\caption{\small{(a) System Model \cite{MTC_Clustered} (b) Association Strategy of a typical NCH.}}
	\label{fig:figure3}
	\end{figure}
Fig. \ref{Cluster_mMTC}  depicts the grouping of MTCDs into multiple disjoint clusters, where each group consists of a single CH member and multiple non-cluster head (NCH) members (as shown in Fig. \ref{System Model}). Moreover, CHs take charge of forwarding the data transmitted by the ordinary nodes in the network. To maximize the energy efficiency of the network, we prioritize the forwarding of NCH member's data through any suitable CH. Failure to associate with all CHs allows a particular NCH to transmit its data directly to the BS. In this work, we assume that each node can participate in the CH election procedure and be elected as a CH or remain an NCH member with probabilities $p$ and $1-p$, respectively. Hence, the average number of CHs and NCHs for a network size of $N$ is equal to ${\lfloor Np \rfloor}$ and $({\lfloor N(1-p) \rfloor})$, respectively, where ${\lfloor x \rfloor}$ indicates the floor function that captures the greatest integer less than or equal to $x$.
We express the disruption of the network by random node removal, mainly caused by physical damage and the energy constraint of sensor nodes.
More specifically, we define the temporal robustness metric $(R_N)$ for a time-varying network as follows:
\begin{align}
R_N =  \frac{\mathbb{E}[N_\text{SuccDisrupted}]}{\mathbb{E}[N_\text{Succ}]}
\label{Robust_def}
\end{align}
The denominator of the robustness metric in (\ref{Robust_def}) represents the mean number of successfully communicating nodes before any disruption in the network. 
Hence, it captures only the natural effect, like communication failure in the network.
On the other hand, in the numerator we evaluate the mean number of successfully communicating nodes after the network is disrupted. 
In the next section, we start by analyzing the denominator of (\ref{Robust_def}) and then focus on the numerator.

\section{Stochastic Analysis of Temporal Robustness Metric}
\subsection{Expected Number of Successfully Communicating Nodes Before the Network Disruption}
To evaluate the number of successfully communicating nodes in the presence of communication failure, we follow the association strategy of a typical NCH member when there are ${\lfloor Np \rfloor}$ CHs in the network, as shown in Fig. \ref{Association Strategy}. 
We assume homogeneity in the network, where all nodes use the same transmitted power, and the received signal power decides the link presence/absence between nodes. A typical NCH, $j$, fails to transmit its data to the BS if both of the following events occur:\\
 \textbf{Event 1}: The $j$th NCH fails to find any suitable CH for its association, or no link exists between the serving CH and the BS. In other words, the $i$th CH will not serve the $j$th NCH if either of the edges between $i$th CH to NCH $j$ or between BS to CH $i$ or both are absent.\\
\textbf{Event 2}: Due to the limited communication range, no direct connection is present between the NCH and the BS.\\ 
The probabilistic analysis of the two events mentioned above is expressed as follows:
Note that there exists no edge from $i$th CH to the $j$th NCH if and only if 
\begin{align}
&P_t\left[\sqrt{(x_j-x_i)^2+(y_j-y_i)^2}\right]^{-\alpha}h_{ji}<P_{th}
\label{No edge CH to NCH}
\end{align}
where $P_t$ is the transmit power of all nodes and  $(x_j, y_j)$, $(x_i, y_i)$ are the coordinates of NCH $j$ and CH $i$, respectively. $\alpha$ denotes the path loss exponent. The $h_{ji}$ represent the power gain of the Rayleigh fading channel between the $i$th CH and $j$th NCH, hence is expressed as exponentially distributed random variables with a mean of 1.
Moreover, $P_{th}$ denotes the received power threshold value that determines the network connectivity. To make the equation compact, from here onward we express  $\sqrt{{(x_j-x_i)^2+(y_j-y_i)^2}}$ as the distance between two vectors, i.e., $|\textbf{r}_j-\textbf{r}_i|$. Hence, (\ref{No edge CH to NCH}) can be rewritten as:
\begin{align}
&P_t|\textbf{r}_j-\textbf{r}_i|^{-\alpha}h_{ji}<P_{th} ~~~~\nonumber \\
& \Rightarrow h_{ji} < \frac{P_{th} |\textbf{r}_j-\textbf{r}_i|^{\alpha} }{P_t}
\label{No connection NCH CH}
\end{align}
Similarly, from the perspective of the downlink transmission, it easily follows that there exists no edge between the BS and the $i$ th CH iff 
\begin{align}
g_{ib} < \frac{P_{th} |\textbf{r}_i|^{\alpha} }{P_B}
\label{No connection CH BS}
\end{align}
where $P_{B}$ is the transmit power of the BS. The $g_{ib}$ represent the power gain of the Rayleigh fading channel between the $i$th CH and the BS. Eq. (\ref{No connection NCH CH}) confirms that the connection probability between the $i$th CH and $j$th NCH is:
$\left(\exp{\left(-\frac{P_{th} {|\textbf{r}_j-\textbf{r}_i|}^{\alpha}}{P_t}\right)}\right)$.
 Similarly, the connection probability between the BS and the $i$th CH from (\ref{No connection CH BS}) can be expressed as:
 $\left(\exp{\left(-\frac{P_{th} |\textbf{r}_i|^{\alpha}}{P_B}\right)}\right)$. So, the $j$th NCH will not find the $i$th CH as a suitable one with the probability of:
\begin{align}
& 1- \exp \left\{-\left(\frac{P_{th} |\textbf{r}_j-\textbf{r}_i|^{\alpha}}{P_t}+ \frac{P_{th} |\textbf{r}_i|^{\alpha}}{P_B} \right)\right\}\nonumber
\end{align}
which is the probability that at least one edge is absent between the two edges mentioned above. 
Thus, the non-association probability of the $j$th NCH with either of the ${\lfloor Np \rfloor}$ CHs in the network is:
\begin{align}
& \prod_{i=1}^{\lfloor Np \rfloor}{\left(1- \exp\left\{-\left(\frac{P_{th} |\textbf{r}_j-\textbf{r}_i|^{\alpha}}{P_t}+ \frac{P_{th} |\textbf{r}_i|^{\alpha}}{P_B} \right)\right\}\right)}\nonumber
\end{align}
where $\{{\textbf{r}_i\}_{i=1}^{\lfloor Np \rfloor}}~\text{and}~{\textbf{r}_j}$~represent the vectors corresponding to the coordinates of all CHs and the $j${th} NCH in the network, respectively. As mentioned earlier, one NCH will attempt to transmit directly to the BS if it fails to associate with any of the CHs in the network. Therefore, the $j$th NCH is not directly connected to the BS iff
\begin{align}
P_B |\textbf{r}_j|^{-\alpha}k_{jb}< P_{th}\nonumber
\end{align}
where $k_{jb}$ represents the power gain of the Rayleigh fading channel between the $j$th NCH and the BS, which confirms that the probability that there is no direct connection between $j$ th NCH and BS is:
$\left(1-\exp{\left(-\frac{P_{th} |\textbf{r}_j|^{\alpha}}{P_B}\right)}\right)$.
So, the communication failure probability of a typical NCH, $j$, or in other words, the probability that the $j$th NCH fails to transmit its data successfully through either way of transmission, can be expressed as:
\begin{align}
\hspace{-2 mm}\mathbb{P}_{FBE} =&\left\{\prod_{i=1}^{\lfloor Np \rfloor}{\left(1- \exp\left\{-\left(\frac{P_{th} |\textbf{r}_j-\textbf{r}_i|^{\alpha}}{P_t}+ \frac{P_{th} |\textbf{r}_i|^{\alpha}}{P_B} \right)\right\}\right)}\right\}\nonumber\\&~~~~~~~~~~~~~~~~\times \left(1-\exp{\left(-\frac{P_{th} |\textbf{r}_j|^{\alpha}}{P_B}\right)}\right)
\label{Fail NCH}
\end{align}
Eq. (\ref{Fail NCH}) is valid for fixed positions of all the CHs and the $j$th NCH. However, in this work, the NCHs and the CHs can be positioned anywhere within the grid dimension of $[-a, a]$. Hence, the mean failure probability of NCH $j$, expecting over all grid positions of both $j$th NCH and all CHs, can be expressed as:
\begin{align}
&\mathbb{P}_{FNCH} =
\mathbb{E}_{\textbf{r}_j,\{\textbf{r}_i\}_{i=1}^{\lfloor Np \rfloor}}\left[\mathbb{P}_{FBE}\right]
\label{Failed_Prob_NCH}
\end{align}
Similarly, the expected successful communication probability of CH $i$, $\mathbb{P}_{SCH}$, over all possible grid positions is:
\begin{align}
\mathbb{P}_{SCH} =\mathbb{E}_{\textbf{r}_i} \left[ \exp{\left(-\frac{P_{th} |r_i|^{\alpha}}{P_B}\right)}\right] \nonumber
\end{align}
Hence, for the given values of CHs (i.e., $N_{CH}$) and NCHs (i.e., $N_{NCH}$), the average number of successfully communicating NCHs and CHs in the network are equal to:
\begin{align}
N_{SNCH} = N_{NCH}(1-\mathbb{P}_{FNCH})\nonumber
~~\&~~
N_{SCH} = N_{CH}\mathbb{P}_{SCH}\nonumber
\end{align}
\begin{figure*}[ht]
 \hrulefill
\begin{align}
\hspace{-6 mm}\mathbb{P}_{FNCH} =
&\mathbb{E}_{\textbf{r}_j,\{\textbf{r}_i\}_{i=1}^{\lfloor Np \rfloor}}\left[\left\{\prod_{i=1}^{\lfloor Np \rfloor}{\left(1- \exp\left\{-\left(\frac{P_{th} |\textbf{r}_j-\textbf{r}_i|^{\alpha}}{P_t}+ \frac{P_{th} |\textbf{r}_i|^{\alpha}}{P_B} \right)\right\}\right)}\right\} \times \left(1-\exp{\left(-\frac{P_{th} |\textbf{r}_j|^{\alpha}}{P_B}\right)}\right)\right]
\label{Prob_NCH}
 \end{align}
\begin{align}
~~~&\approx\mathbb{E}_{\textbf{r}_j,\{\textbf{r}_i\}_{i=1}^{\lfloor Np \rfloor}}\left[\left\{\prod_{i=1}^{\lfloor Np \rfloor}{\left(1- \exp\left\{-\left(\frac{P_{th} |\textbf{r}_j-\textbf{r}_i|^{\alpha}}{P_t}+ \frac{P_{th} |\textbf{r}_i|^{\alpha}}{P_B} \right)\right\}\right)}\right\}\right] \times \mathbb{E}_{\textbf{r}_j}\left[\left(1-\exp{\left(-\frac{P_{th} |\textbf{r}_j|^{\alpha}}{P_B}\right)}\right)\right]\nonumber
 \end{align}
 \begin{align}
~~~&\approx \prod_{i=1}^{\lfloor Np \rfloor}\mathbb{E}_{\textbf{r}_j,\{\textbf{r}_i\}_{i=1}^{\lfloor Np \rfloor}}\left[\left\{{\left(1- \exp\left\{-\left(\frac{P_{th} |\textbf{r}_j-\textbf{r}_i|^{\alpha}}{P_t}+ \frac{P_{th} |\textbf{r}_i|^{\alpha}}{P_B} \right)\right\}\right)}\right\}\right] \times \mathbb{E}_{\textbf{r}_j}\left[\left(1-\exp{\left(-\frac{P_{th} |\textbf{r}_j|^{\alpha}}{P_B}\right)}\right)\right]\nonumber
 \end{align}
   \begin{align}
\hspace{3 mm} & = \left\{\mathbb{E}_{\textbf{r}_j,\{\textbf{r}_i\}_{i=1}^{\lfloor Np \rfloor}}\left[{\left(1- \exp\left\{-\underbrace{\left(\frac{P_{th} |\textbf{r}_j-\textbf{r}_i|^{\alpha}}{P_t}+ \frac{P_{th} |\textbf{r}_i|^{\alpha}}{P_B} \right)}_{z_1+z_2}\right\}\right)}\right]\right\}^{\lfloor Np \rfloor} \times \mathbb{E}_{\textbf{r}_j}\left[1-\exp{\left(-\underbrace{\frac{P_{th} |\textbf{r}_j|^{\alpha}}{P_B}}_{z_{DBS}}\right)}\right]
\label{approximated}
\end{align}
   \begin{align}
\hspace{6 mm}&=\left[{\frac{1}{(2a)^4}}\int_{-a}^a\int_{-a}^a \int_{-a}^a\int_{-a}^a  \left(1-\exp{\left(-(z_1+z_2)\right)}\right)dxdydx_1dy_1\right]^{\lfloor Np \rfloor} \left[\frac{1}{(2a)^2} \int_{-a}^a\int_{-a}^a \left(1-\exp{\left(-{z_{DBS}}\right)}\right) dxdy \right]
 \label{Expand_Eq}
 \end{align}
\hrulefill
\end{figure*}
So, before any node removal in the network, the total number of successfully communicating nodes in the network is equal to   
\begin{align}
N_{SPA} = N_{NCH}(1-\mathbb{P}_{FNCH}) + N_{CH}\mathbb{P}_{SCH}
\label{Pre_Attack_Total_Success}
\end{align}
Note that (\ref{Pre_Attack_Total_Success}) is valid for a fixed number of CHs and NCHs in the network. However, in this work we allow the number of CHs and NCHs to vary between $0$ and $N$. Moreover, as the numbers of CHs and NCHs are dependent on each other, i.e., $N_{CH} = N- N_{NCH}$, the expectation over any one of these two is sufficient to proceed further.
The mean number of successfully communicating nodes expecting over all possible values of CHs is:
\begin{align}
N_{ESPA} =    &\mathbb{E}_{N_{CH}}\left[ N_{NCH}(1-\mathbb{P}_{FNCH}) + N_{CH}\mathbb{P}_{SCH}\right]\nonumber \\ =  & \lfloor N(1-p) \rfloor(1-\mathbb{P}_{FNCH}) + \lfloor Np \rfloor\mathbb{P}_{SCH}
\label{Pre_Attack Succ}
\end{align}
 Eq. ({\ref{Pre_Attack Succ}}) represents the average number of successful nodes in the presence of communication failure only. In the next section, we will capture the effect of random node failure in the network.
 \subsection{Expected Number of Successfully Communicating Nodes in the Disrupted Network}
 To investigate the impact of the random node failure on the number of successfully communicating nodes, let us assume that at a particular instant of time, the number of failing nodes in the network is $K$, where $1\leq K \leq N$ and $N= N_{CH} + N_{NCH}$. Suppose that among the $K$ failing nodes, there are $R$ CHs and $K-R$ NCHs removed from the network. Hence, the remaining number of CHs and NCHs after $K$ nodes removal are $N_{CH}- R$ and $N_{NCH}-K+R$, respectively. 
 By following (\ref{Pre_Attack_Total_Success}), the number of successfully communicating nodes after the $K$ nodes are removed from the network is equal to: 
 \begin{align}
\mathbb{L}_0(K,R,N_{CH}) = &[\{(N-N_{CH})- (K-R)\}(1-\mathbb{P}_{FNCH})]\nonumber \\~~~~& + (N_{CH}- R)\mathbb{P}_{SCH}
\label{Succ Nodes Fixed}
\end{align}
Eq. (\ref{Succ Nodes Fixed}) is valid for fixed values of $K,R,N_{CH}$. However, the stochastic analysis should express the mean robustness for all possible values of $K, R, N_{CH}$, which we now look into. Note that for a given $K$, the value of $R$ can vary within a specific range, which is determined from $0 \leq R \leq K$, $0 \leq R \leq N_{CH}$ and  $0 \leq K-R \leq N- N_{CH}$. From these conditions, one can easily derive that $\underbrace{\text{max}\{0,K-N+N_{CH}\}}_{R_{min}}\leq  R \leq \underbrace{\text{min}\{K,N_{CH}}_{R_{max}}\}$. 
Hence, the expected  number of successful nodes for all possible values of the number of removed CHs, $R$, is expressed as:
\begin{align}
\mathbb{L}_1(K,N_{CH}) = &~~\mathbb{E}_{R}
\left[\mathbb{L}_0(K,R,N_{CH})\right] \nonumber\\ = & \sum_{R= R_{min}}^{R_{max}}\mathbb{L}_0(K,R,N_{CH})\mathbb{P}_{FCH}(R,K,N_{CH})
\label{Succ over R}
\end{align}
$\mathbb{P}_{FCH}(R, K, N_{CH})$ represents the failure probability of $R$ CHs among the $K$ removed nodes, and can therefore be expressed as:
\begin{align}
\mathbb{P}_{FCH} (R,K,N_{CH}) = \frac{\binom {N_{CH}}R\binom {N-N_{CH}}{K-R}}{\binom NK}\nonumber
\end{align}
 One can easily derive that the number of failing nodes is uniformly distributed and can take any value in $\kappa = \{1,2,3,.., N\}$. Hence, expecting over all possible numbers of removed nodes, the number of successfully communicating nodes becomes:
 \begin{align}
\mathbb{L}_2(N_{CH})
=\mathbb{E}_{K}
\left[\mathbb{L}_1(K,N_{CH})\right]\ = \frac{1}{N}\sum_{K =1}^{N}\mathbb{L}_1(K,N_{CH}) \nonumber
\end{align}
where $\mathbb{P}_{K}(N) = \frac{1}{N}$ is the probability mass function (PMF) of the number of removed nodes in the network. Similarly, the expectation over the possible number of CHs gives:
\begin{align}
\mathbb{L}_3 = \sum_{N_{CH} =0}^{N}\underbrace{\mathbb{L}_2({N_{CH}})}_{f(N_{CH})} \mathbb{P}_{N_{CH}}(N,p)
\label{L3}
\end{align}
where $p$ is the CH probability and 
$\mathbb{P}_{N_{CH}}(N,p)$ is the PMF of the number of CHs, $N_{CH}$. As an individual node can be elected either as a CH or as an NCH member, the PMF of the number of CHs follows the binomial distribution, which can be expressed as:
\begin{align}
 \mathbb{P}_{N_{CH}}(N,p) = \binom{N}{N_{CH}}p^{N_{CH}}(1-p)^{N-N_{CH}}\nonumber
 \end{align}
 where we recall that $p$ represents the probability of a node being elected as a CH.
Now, by following Eqs. (\ref{Robust_def}), (\ref{Pre_Attack Succ}), and  (\ref{L3}), the temporal robustness metric can be expressed as:
\begin{align}
\hspace{-6 mm}R_N  = \frac{\mathbb{L}_3}{N_{ESPA}}
\label{Robustness Analytical}
 \end{align}
Note that the first factor of the exact expression of temporal robustness expressed in (\ref{Prob_NCH}) exhibits the form of $\mathbb{E}[f(\textbf{r}_j,\textbf{r}_1),....,f(\textbf{r}_j,\textbf{r}_{\lfloor Np \rfloor})]$, where the presence of the common variable, $\textbf{r}_j$ in all the terms in the product form makes their separation difficult. The dependency on this variable, $\textbf{r}_j$, complicates the evaluation of (\ref{Prob_NCH}) for a higher number of CHs. Therefore, to make the equation simpler, we use two levels of approximation:\\
\textbf{Approximation 1:}
The above discussion confirms that (\ref{Robustness Analytical}) becomes intractable for a large number of CHs in the network. However, to simplify the equation,  (\ref{Prob_NCH}) is approximated as
\begin{align}
\hspace{-1 mm}
\mathbb{E}[f(\textbf{r}_j,\textbf{r}_1),..,f(\textbf{r}_j,\textbf{r}_{\lfloor Np \rfloor})] \approx \mathbb{E}[f(\textbf{r}_j,\textbf{r}_1)]\times..\times \mathbb{E}[f(\textbf{r}_j,\textbf{r}_{\lfloor Np \rfloor})]
\label{Approximated Form1}
\end{align}
Note that all the ${\lfloor Np \rfloor}$ elements of the first factor in (\ref{Prob_NCH}) end up having the same expected values due to positioning within the same grid range from $-a~ \text{to}~a$. Hence, the above expression is approximated in (\ref{approximated}).\\ 
\textbf{Approximation 2:}
Note that (\ref{L3}) can be represented in the form of 
$\mathbb{E}_{N_{CH}}\left[ f({N}_{CH})\right]$. 
By using the following  approximation $\mathbb{E}_{N_{CH}}\left[ f({N}_{CH})\right] \approx  f(\mathbb{E}_{N_{CH}}\left[{N}_{CH}\right]) = f({\lfloor Np \rfloor})$ (as we define earlier that the $\mathbb{E}[N_{CH}] = \lfloor Np \rfloor$), (\ref{L3}) can be approximated as:
\begin{align}
\mathbb{L}_3 \approx \mathbb{L}_2({N_{CH}=\lfloor Np \rfloor})\nonumber
\end{align}
\section{Numerical Investigation}
This section investigates the effectiveness of the proposed analytical model described in section III.  
More specifically, to corroborate the exactness of the analytical expression derived in this work, we here compare the analytical results with the ones we obtain from the numerical investigation.  
The system parameter values used in this investigation are listed in Table I.
Our numerical investigation is performed for the time switching graph over 10,000 realizations corresponding to different node positions, the number of failing nodes, all possible numbers of CHs in the network, and the number of CHs removed due to network disruption. Moreover, we assume a node density value in line with the current requirement of supporting one million devices per $\text{km}^2$ for mMTC services. We assume that all nodes can participate in the CH election procedure, where they generate random numbers between [0,1] and elect themselves as CH members only if the outcome is less than  $p$.
Once the election procedure is over, all NCHs prefer to send their data through a suitable CH to maximize the networks' energy efficiency. If an NCH member fails to associate with all the available ${\lfloor Np \rfloor}$ CHs, only then it attempts to transmit to the BS directly. Note that, unlike an infrastructure network, the radio link in a communication network ensures an effortless reassociation to the network connectivity. Hence, we allow the network reconnection after a disruption occurs. Moreover, we consider a downlink mMTC network that enables the BS to operate as a transmitter for determining the connectivity between BS and a node. 
\begin{figure*}
	\centering
	\begin{minipage}{0.3\linewidth}
	 \vspace{0.3cm}
	    \hspace{-0.2cm}
	    \includegraphics[width=6.1cm]{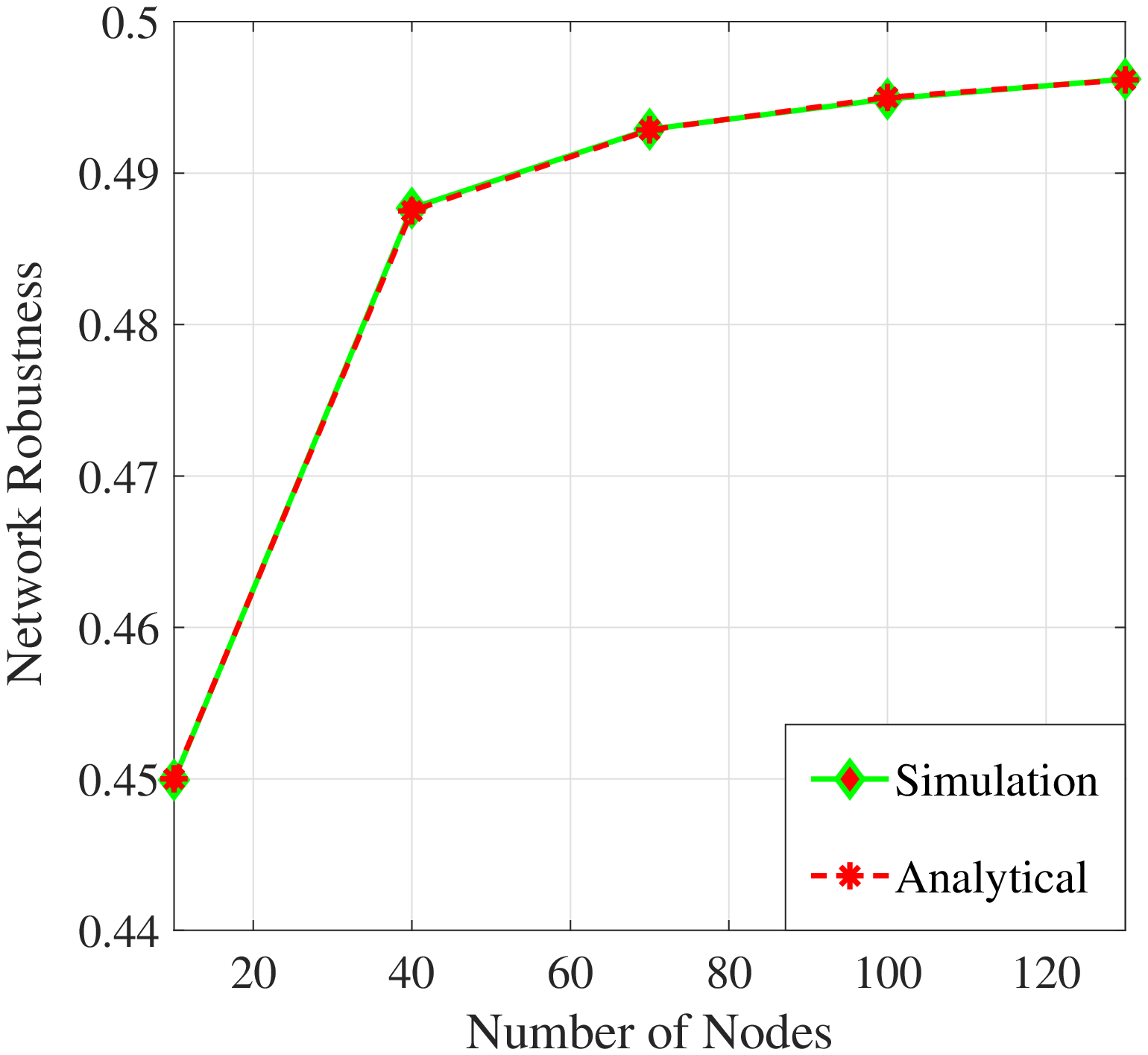}
		\caption{Comparative study of network robustness as a function of network size.}
		 \label{Comparative_Study}
	\end{minipage}%
	\hspace{0.3cm}
	\begin{minipage}{0.33\linewidth}
	    \vspace{0.15cm}
	    \hspace{-0.3cm}
		\includegraphics[width=6.0cm]{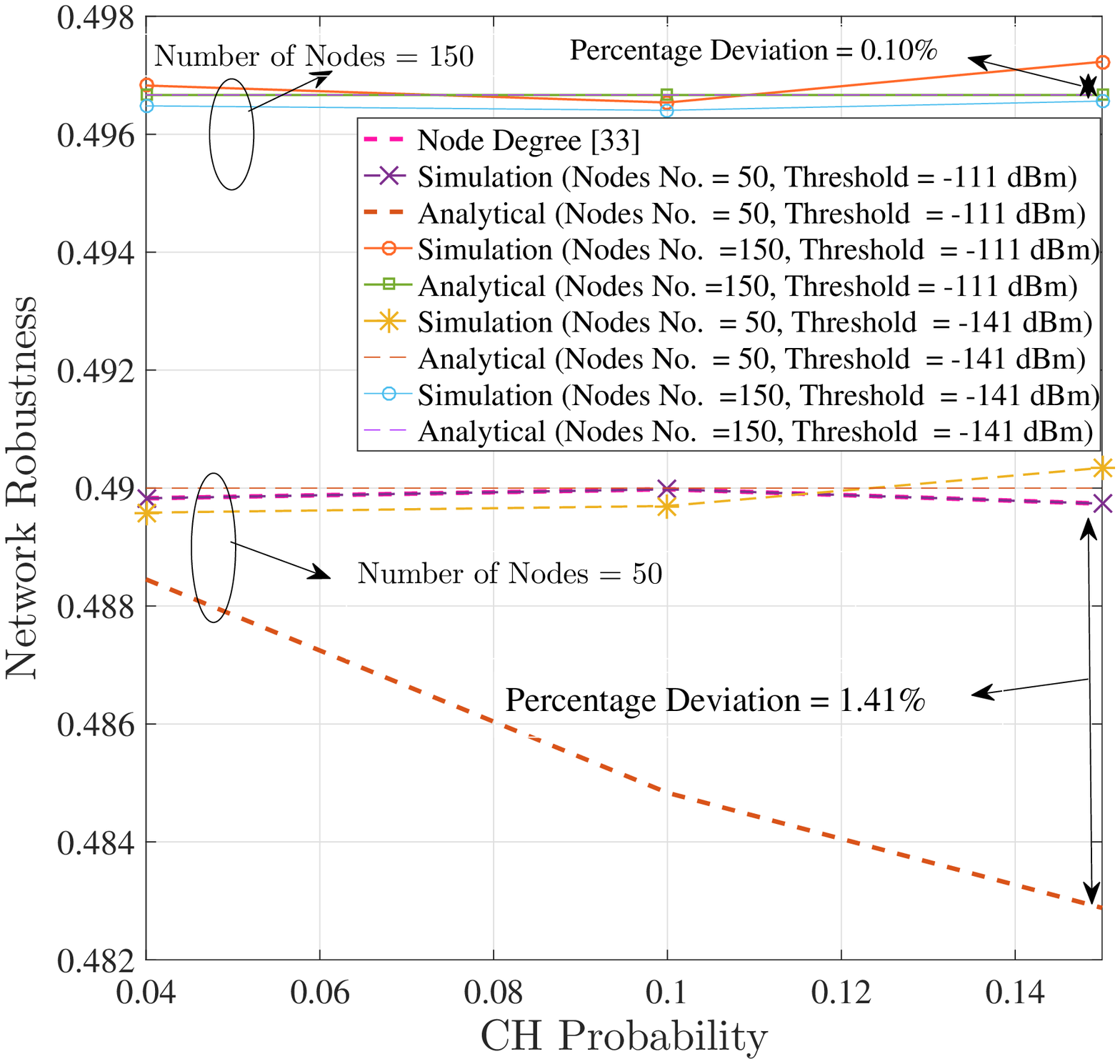}
		\caption{Network robustness comparison as a function of CH probability for different network sizes and power threshold values.}
		 \label{Comparative_Diff_CH}
	\end{minipage}
	\hspace{0.3 cm}
	\begin{minipage}{0.3\linewidth}
	 \vspace{1.2cm}
	   \hspace{-0.4cm}
		\includegraphics[width=6.2cm]{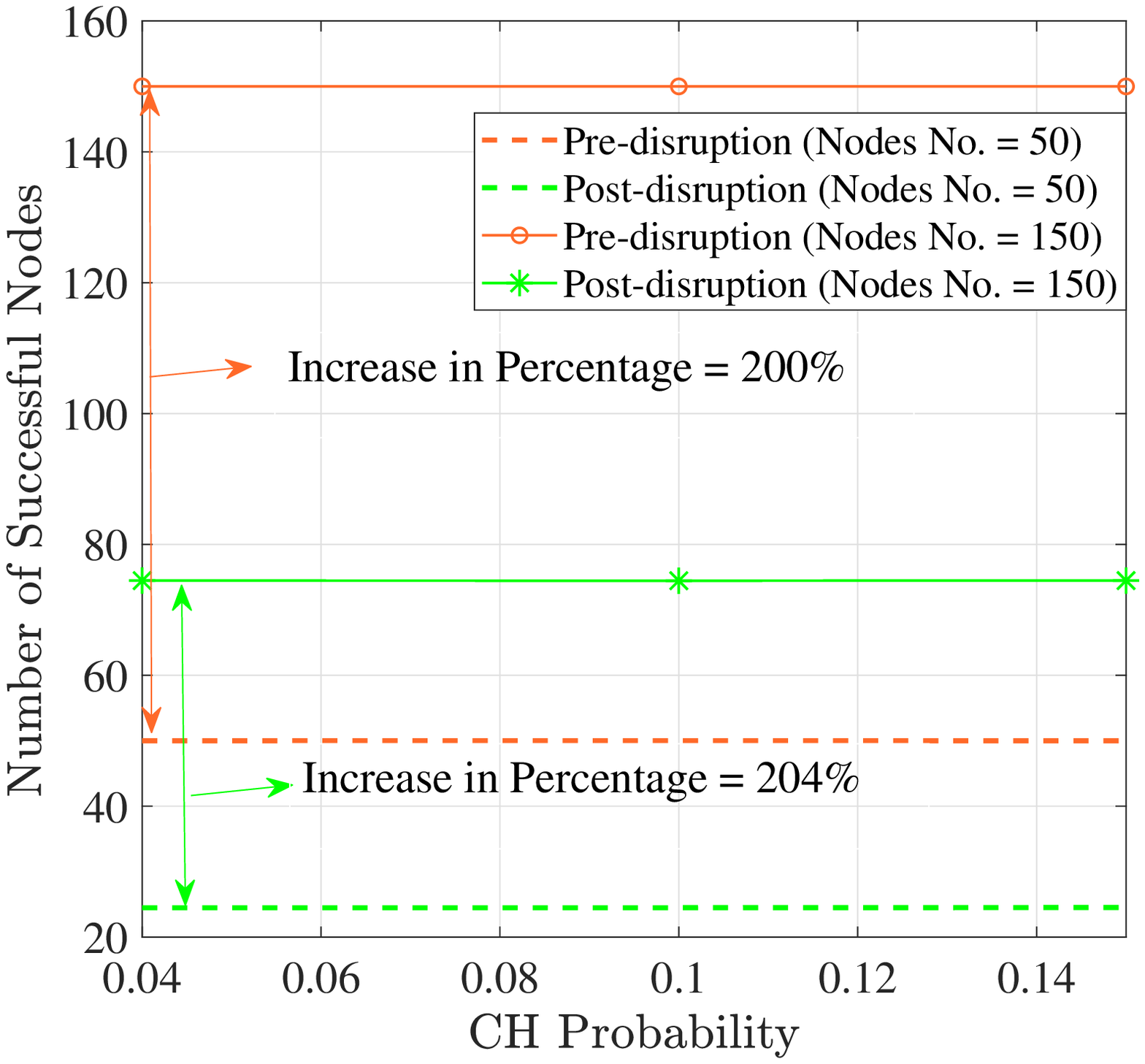}
		\caption{The number of successfully communicating nodes (pre- and post-node removal) as a function of CH probability for different network sizes for power threshold value = -111dBm.}
		\label{Successful_Nodes}
	\end{minipage}
	\end{figure*}
\begin{table}[H]
     \centering
     \begin{tabular}{|p{15mm}|p{18mm}|p{23mm}|p{10mm}|} \hline
	\textbf{Parameter} &\textbf{Value} &\textbf{Parameter} &\textbf{Value}\\\hline
	Number of Nodes & 150 & Node Density & 1 Node/$m^2$\\\hline
	$P_{th}$ & -111 dBm, -141 dBm \cite{Tx_power_mMTC}  & $P_B$ & 46 dBm\\\hline
	$P_{t}$ & 23 dBm \cite{Tx_power_mMTC}  & Number of Iterations &10,000\\\hline
    \end{tabular}
    \caption{Simulation Parameters.}
    \end{table}
Fig. \ref{Comparative_Study} performs a comparative study between the analytical and the numerical investigation. Note that in Fig. \ref{Comparative_Study}, we compare the result of the exact form of the network robustness (i.e., Eq. (\ref{Prob_NCH})) with the numerical one. The overlapping curves corresponding to the exact form of analytical expression and numerical investigation validate the derived analytical framework. As explained earlier, to
circumvent the intractability of (\ref{Prob_NCH}), we present a comparatively
low complexity two-level approximated form of the exact
expression in (\ref{approximated}), which we have exploited in the below section for further investigation.
 
 We now evaluate the trend of the  network robustness and the number of successful nodes as a function of various system parameters.  By considering the quantity defined in \cite{Nature_TR}, we express the mean node degree $\left(\langle K \rangle \right)$ of a time-varying switching graph as:
    \vspace{-2mm}
    \begin{align}
    &K(t) = \frac{1}{N}\sum_{i=1}^{N}k_{i}(t); ~~~\text{and}~~ \langle K \rangle =  \frac{1}{T}\sum_{t=1}^{T}K(t) 
    \label{Average Degree Simulation}
\end{align} 
 where $t =\{1,2,...,T\}$ is the collection of the observation time instants and $k_{i}(t)$ represents the in-degree of the $i$th node at the $t$th instant of time. We compare the numerical investigation result with the network robustness metric described in  (\ref{Average Degree Simulation}). The overlapping curves in Fig. \ref{Comparative_Diff_CH} confirm the correlation of 1 between the temporal robustness value obtained from numerical investigation and the one achieved from \cite{Nature_TR}, thereby establishing a direct relation between the two.
 In addition, Fig.  \ref{Comparative_Diff_CH} compares the results of the numerical investigation with the approximated analytical framework as a function of CH probability for different network sizes. Moreover, this figure exhibits the influence of the power threshold value on the parametric trend of the network robustness metric, which confirms a gap between the analytical and numerical investigation smaller than 1.5\%, even in the presence of a higher value chosen from the realistic range of the power threshold. This gap becomes even more negligible (0.10\%) for a comparatively lower power threshold value of -141 dBm.
 This minimal performance gap in Fig.  \ref{Comparative_Diff_CH} allows us to use the low complexity approximated form of the analytical expression for further investigation. Furthermore, Fig. \ref{Comparative_Diff_CH} demonstrates that with an increase in the number of CHs, the gap between the simulation and the approximated analytical result increases, especially for a higher power threshold value (i.e., -111 dBm). This observation can be justified as follows:
 Note that (\ref{approximated}) is the approximation of (\ref{Prob_NCH}), which uses the form stated in (\ref{Approximated Form1}).
With an increase in the number of CHs (i.e.,${\lfloor Np \rfloor}$), an approximation in the first factor of (\ref{Prob_NCH}) in the form of (\ref{Approximated Form1}),
introduces an additional approximation gap in (\ref{approximated}), which in turn increases the difference between the analytical and numerical investigation for a large number of CHs in the system. Furthermore, this figure exhibits the variation of network robustness as a function of the CH probability.
We observe a negligible variation in the robustness metric value for a wide range of realistic power threshold values as a function of the number of CHs present in the network. The below discussion justifies the observed parametric trend.

Note that the robustness metric in (\ref{Robustness Analytical}) is defined as the ratio between the number of successful nodes after and before the network disruption. Hence, depending on the nature of these two factors,
the parametric trend of the network robustness metric varies with the system parameters. To justify the above-mentioned observed trend in Fig.  \ref{Comparative_Diff_CH}, Fig. \ref{Successful_Nodes} investigates the number of successful nodes before and after network disruption as a function of the number of CHs in the network. The comparable increment of both the factors (i.e., the pre- and post-disruption number of successful nodes) introduces a negligible variance in the network robustness for the different CH probability values. This observation is true for a wide range of realistic power threshold values (where the practical range of power threshold lies between -84 dBm to -141 dBm \cite{Tx_power_mMTC}).
\begin{figure}
\hspace{-3 mm}
\subfigure[Percentage of failing nodes due to the network disruption as a function of CH probability for different network sizes and power threshold values.]{\includegraphics[scale=0.2]{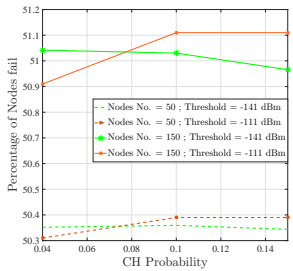}}\quad
\subfigure[Percentage of the CHs failed due to the network disruption as a function of CH probability for different power threshold values and for network sizes =150.]{\includegraphics[scale=0.235]{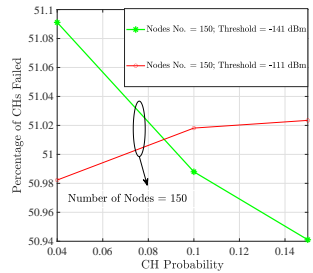}}
\caption{Impact study of network disruption on network nodes.}
  \label{Disruption_Impact}
\end{figure}
For an in depth analysis of the observed trend in Fig. \ref{Comparative_Diff_CH}, Fig.   \ref{Disruption_Impact} evaluates the impact of the network disruption as a function of number of CHs for various network sizes and for different values of the power threshold. Fig.  \ref{Disruption_Impact}(a) confirms an increasing trend of the failing nodes as a function of the number of CHs only when the power threshold value is comparatively high (i.e., -111 dBm). In particular, we express the percentage of failing nodes as the ratio between the number of failing nodes due to the network disruption and the number of successful nodes before network disruption.
In contrast, the percentage of failing nodes exhibits a decreasing trend as a function of number of CHs for a lower power threshold value of  -141 dBm.  Fig.  \ref{Disruption_Impact}(b) justifies the trend of Fig. \ref{Disruption_Impact}(a) as it exhibits a similar trend of increasing and decreasing nature of the percentage of failing CHs, due to the network disruption as a function of CH probability in the presence of a power threshold value of -111 dBm and -141 dBm, respectively. More specifically, the percentage of failing CHs is measured by the ratio between the number of failing CHs due to network disruption and the number of successful CHs under the network's standard operation. The observed trends can be justified as follows: an increase in the CH probability indicates more CHs in the network, thereby decreasing the number of NCHs in the network as the sum of CHs and NCHs is equal to $N$. Note that the network connectivity only in the presence of a high received power demands a reliable channel between one device and BS for the successful transmission. An NCH member's operation enables a node to either forward its data through a CH member or transmit its data directly if no suitable CH is found for relaying the data. In contrast, only the presence of a direct link between the CH to the BS can ensure the successful communication of a CH member. Hence, with increased CHs, the extra benefit of using relay nodes decreases in the system as all CHs need to send their data directly to the BS. The lack of channel diversity due to the strict requirement of direct connection to the BS reduces the possibility of experiencing reliable channels by each CH member. As a result, the possibility of CH failure increases with increasing number of CHs in the presence of a high connectivity threshold. In contrast, a low power threshold value requires much less received power to guarantee the network connectivity, which can be easily ensured by direct transmission between one node and the BS without requiring the support of any relay node. In addition, the random placements of NCH nodes over the pre-specified grid position restrict the power-constrained NCH from reaching any CH in the presence of fewer CHs in the network. However, with an increase in the number of CHs, the possibility that one typical NCH will opt for an intermediate node to forward its data also increases. Thus the higher possibility of direct communication and the added benefit of exploiting an intermediate node prevent the number of failing nodes from increasing further for a low power threshold, which explains the decreasing trend of the percentage of failing CHs as a function of the CH probability in the system. Hence, in the presence of a fixed number of nodes in the network, the percentage of failing nodes directly follows the trends of failing CHs percentage for all power threshold values.

In addition, from Fig. \ref{Comparative_Diff_CH} we can observe that with an increase in network size, the network robustness enhances, where a high robustness value for large networks indicates the strong potential of any network to continue its regular operation by combating any external disturbance and challenges. The observed parametric trend can be explained by following the similar argument stated above. In detail, an increase in the network size has a more positive influence on the number of successful nodes after the disruption, than that on the number of successfully communicating nodes before the network is disrupted. More precisely, a comparatively
higher increment in the post-disruption number of successful nodes (i.e., 204\%) than that of the pre-disruption successful nodes (i.e., 200\%) with increasing network size in Fig.  \ref{Successful_Nodes} justifies the rising tendency of network robustness metric as a function of network size in Fig. \ref{Comparative_Diff_CH}. Hence, Figs.  \ref{Successful_Nodes}-\ref{Disruption_Impact} combinedly justify both the increasing trend and the negligible variation in the network robustness in the presence of large network size and for different power thresholds values, respectively.
\begin{figure}
    \centering
    \includegraphics[width=0.75\linewidth]{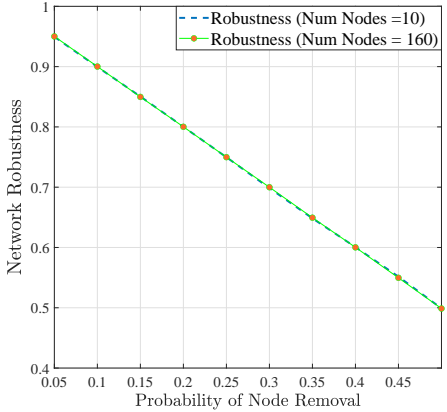}
     \caption{Network robustness as a function of node failure probability.}
     \label{Impact_Node_Failure}
     \end{figure}
Finally, we investigate the impact of the node removal on the temporal robustness factor. The decreasing trend of the robustness for an increasing node failure probability, implies the rapid degradation in a network's capability to combat disruption and challenges. 
The above finding is aligned with the results in \cite{Scellato_1}.
\vspace{-2 mm}
  \subsection{Computational Complexity}
The below section compares the computational complexities of a few well-established network robustness metrics with the one we propose. In this context, all prior robustness metrics, i.e.,  super connectivity, toughness, scattering number, tenacity, expansion  parameter, isoperimetric number, min-cut, and the $k$-connectivity evaluate the network robustness in an NP-time, hence, are unsuitable for large networks.
Therefore, to clearly define the advantage of our proposed robustness metric, we now provide a step-by-step time complexity calculation. As mentioned earlier, we perform a stochastic analysis of the temporal robustness metric over all possible node positions in grid, the number of CHs in the network, all numbers of removed nodes, and corresponding feasible range of removed CHs caused by network disruption. Note that the NCH failure probability expecting over all the node positions can be expressed by (\ref{Expand_Eq}), which has constant time complexity. However, the dependency on the number of CHs in the network, number of nodes removal, and the feasible range of the removed CHs can be characterized by the nested \textit{for} loops with the maximum number of iterations of $(N+1)$, $N$, and $(K+1)$, respectively. Note that the maximum number of CHs that can be removed from the network, $K$, is equal to $N$. The above fact confirms that the time complexity of the temporal robustness metric evaluation is $(N+1) \times N\times (N+1)~=~ N^3+2N^2+N$. Hence, the worst-case time complexity of our proposed scheme is $O(N^3)$, which is the same as the algebraic connectivity described in \cite{Algebric_Connectivity_Survey}.
Moreover, the connectivity evaluation in a network requires $O(M^2N^2)$ time complexity \cite{Natural_Connectivity}. In addition, as shown in Approximation 2, the two-level approximated form stated in (\ref{approximated}) excludes the requirement of the \textit{for} loop associated with the number of CHs in the network, thereby experiencing the time complexity of $O(N^2)$.  The low complexity of our proposed metric makes it a particularly promising candidate for temporal robustness measurement, especially for large networks.
\section{Conclusion}
 Robust connectivity is one of the essential requirements of communication networks. The demand for a reliable connection in the latest 5G deployment scenarios led to recent investigations to revisit the topic of robustness analysis. Unlike fixed networks, wireless communication networks experience both the random fluctuation and uncertainty of the links and nodes present in the network, limiting the application of existing works and motivating the necessity for further investigations in this domain. The lack of complete analysis in terms of temporal robustness inspired us to perform a stochastic analysis of temporal robustness in mMTC networks in this article. The comparative study with the numerical investigation confirms the exactness of our proposed analytical framework.
Furthermore, the encapsulation of the communication range effect and the network's temporal variation enables the proposed metric to quantify the temporal robustness in a realistic communication network. Besides, the low complexity and minimal gap with respect to the exact analytical expression confirm the suitability of the introduced two-level approximated form as a candidate low-complexity temporal robustness metric, proving particularly useful for large-scale deployments. 
\vspace{-2 mm}
\bibliographystyle{IEEEtran}
\bibliography{Ref}

\end{document}